\documentclass[%
reprint,
amsmath,amssymb,
aps,
]{revtex4-1}

\usepackage{graphicx}
\usepackage{dcolumn}
\usepackage{bm}
\usepackage[mathlines]{lineno}
\usepackage{natbib}

\begin{document}
	
	\preprint{APS/123-QED}
	
	\title{Implications of single field inflation in general cosmological scenarios on the nature of dark energy given the swampland conjectures}

	\author{Oem Trivedi}%
	\email{oem.t@ahduni.edu.in}
	\affiliation{%
		School of Arts and Sciences, Ahmedabad University,Ahmedabad 380009,India}%

	\date{\today}
	
	\begin{abstract}
		Swampland Conjectures have attracted quite some interest in the Cosmological Community. They have been shown to have wide ranging implications , like Constraints on Inflationary Models, Primordial Black Holes etc. to name a few.  A particularly revealing insight on dark energy also shows that one can have the dark energy equation of state for a quintessence scenario to be signficantly different than -1 after one takes into account the refined dS conjecture.  Another interesting issue with the swampland conjectures is that they have been shown to be incompatible with single field inflationary models in GR based cosmology. In our previous work we have, however, showed that single field inflationary models are quite compatible with swampland conjectures in their usual string theoretic form in a large class of modified cosmological scenarios. Building on that work,  we now show that in modified cosmological scenarios where the early universe expansion was driven by single field inflation , one can have the dark energy equation of state to be significantly different from -1 even if we just take into account the original dS conjecture, let alone the refined form of that.  We thereby show that one does not need to apply a step function approach towards inflation in order to have an observable distinction between constant and non constant dark energy models in the context of the swampland conjectures.
	\end{abstract}
	
	\maketitle
	
	\section{Introduction}
	Inflationary Cosmology has achieved a tremendous amount of success in describing the Very Early Universe \cite{sato1981first,baumann2009tasi,guth1981inflationary,linde1983chaotic,linde1995quantum}.A huge amount of inflationary models \cite{martin2014encyclopaedia,wands2008multiple,berera1995warm,ashtekar2010loop,langlois2008perturbations,golovnev2008vector,setare2013warm,kanti2015gauss,dvali1999brane,alexander2013chern} have been shown to be very consistent with the most recent datasets on the early universe \cite{aghanim2018planck,akrami2020planck1,akrami2018planck,aghanim2020planck,martin2014best}.The consistency of Inflation with observational data for large amount of diversely motivated models is a really eye catching property of Inflationary Cosmology. Besides explaining various problems of the early universe in  standard big bang cosmology, inflation has also provided a way to understand the current dark energy epoch of the universe. The idea of Quintessential Inflation has garnered quite a lot of interest recently and in this, the Inflaton field itself becomes the cause of the accelerated expansion of the universe today \cite{peebles1999quintessential,caldwell1998cosmological,tsujikawa2013quintessence,zlatev1999quintessence,bento2002generalized,dimopoulos2019warm,benisty2020quintessential,dutta2008hilltop,chiba2009slow}. The idea of Quintessence, like Inflation, has been thoroughly studied in Modified Cosmologies as well \cite{gardner2007braneworld,sami2004quintessential,koivisto2007gauss,perrotta1999extended,van2017gauss,benisty2020quintessential1}.
	\\
	In recent years, a quest for a unified theory of everything has garnered a lot of attention in high energy physics. Probably the most well known of these "theory of everything" candidates, is String Theory \cite{green1981supersymmetrical,green1982supersymmetric,green1982supersymmetric1,gubser1998gauge,seiberg1999string,garfinkle1991charged,susskind2003anthropic,polchinski1998string,lust1989lectures}. It is one of the most viable approaches for a theory of everything for many (if not the most viable), hence one would expect that String theory has some implications for Cosmology and could also hence allow us to understand both the very early universe and it's subsequent evolution in more detail. So it's no surprise that a substantial amount of work has been done on studying the cosmological implications of String theory \cite{mcallister2008string,gasperini1994dilaton,gasperini2003pre,gasperini2007elements,tseytlin1992elements,krori1990some,heckman2019f,heckman2019pixelated}. One curious feature of this theory is the extremely high amount of vacua states that is posits, going around $ \mathcal{O} (10^{500}) $ and this constitutes the so called "String Landscape". But then, a fundamental question that arises is how can one distinguish between low energy effective field theories which are consistent with String theory and those which are not. In a bid to understand this question in more detail, Vafa in \cite{vafa2005string} gave the term "Swampland" to refer to the set of all low energy EFT's which are inconsistent with the String paradigm. As String theory is (as a consequence of being a theory of everything) seen as a viable quantum gravitational setup by string theorists, then this would go on to mean that the theories in the swampland are also not viable with a self consistent theory of Quantum gravity. Further, to know as to if a given low energy EFT belongs in the swampland, a number of field theoretic criterion known as the "Swampland conjectures " have been proposed in recent years \cite{obied2018sitter,ooguri2016non,mcnamara2019cobordism,bedroya2019trans,garg2019bounds}.The prominent Swampland Conejctures which gathered immediate interest in the context of Cosmology were :
	\\
	\\
	$ 1 $ : Swampland Distance Conjecture (SDC) : This conjecture limits the field space of validity of any effective field theory \cite{ooguri2016non} . This sets a maximum range traversable by the scalar fields in an EFT as \begin{equation}
	\Delta \phi \leq d \sim \mathcal{O} (1)
	\end{equation} 
	where we are working in the Planck Units $ m_{p} = 1 $ where $ m_{p} $ is the reduced Planck's constant, d is some constant of $ \mathcal{O} (1) $ , and $\phi$ is the Scalar Field of the EFT. \\ \\
	$ 2 $ Swampland De Sitter Conjecture (SDSC) : This Conjecture states that it is not possible to create dS Vacua in String Theory \cite{obied2018sitter}. The conjecture is a result of the observation that it has been very hard to generate dS Vacua in String Theory \cite{dasgupta2019sitter,danielsson2018if}( While it has been shown that creating dS Vacua in String Theory is possible in some schemes ,like the KKLT Construction \cite{kachru2003sitter}). The Conjecture sets a lower bound on the gradient of Scalar Potentials in an EFT , \begin{equation}
	\frac{| V^{\prime} |}{V} \geq c \sim \mathcal{O} (1)
	\end{equation} 
	where c is some constant of $ \mathcal{O} (1) $ , and V is the scalar Field Potential. Another " refined " form of the Swampland De Sitter Conjecutre (RSDSC) places constraints on the Hessian of the Scalar Potential (a finding which first appeared in \cite{garg2019bounds} and later in \cite{ooguri2019distance} ). Expressed in $ m_{p} = 1 $ units, it reads \begin{equation}
	\frac{V^{\prime \prime}}{V} \leq - c^{\prime} \sim \mathcal{O} (1)
	\end{equation}
	where $ c^{\prime} $ is again some constant of Order 1. 
	\\
	\\
	The conjecture has had quite fascinating implications for single field inflation, particularly those regimes in a General Relativistic Cosmology. It was shown in \cite{kinney2019zoo} that the swampland conjectures Eq. (1-3) are not consistent with the cosmological data on single field inflation in a GR based cosmology and for them to be viable with this form of Inflation, the string theory- motivated definitions of the conjectures would have to change. A lot of work has since then been done to understand the issues of the distance and de sitter conjectures with single field Inflation \cite{trivedi2020swampland,geng2020potential,scalisi2019swampland,scalisi2020inflation,ashoorioon2019rescuing}.Besides these conjectures, another recently proposed swampland conjecture by the name of the " Trans Planckian Censorship Conjecture(TCC)" \cite{bedroya2019trans} , implies that single field Inflation in a GR Based Cosmology would have to be severely fine tuned in order for it to not lie in the swampland \cite{bedroya2020trans} . If Inflation is plagued with severe fine tuning problems itself, which were the kind of issues in standard big bang cosmology which prompted work on Inflation in the fist case, then it is certainly a very dire situation for Inflationary Cosmology keeping in mind these conjectures. A lot of work has been done in order to understand the issues of single field GR based inflation and the TCC in more detail \cite{brahma2020trans,brandenberger2020strengthening,dhuria2019trans,kamali2020relaxing,bernardo2020trans,mizuno2020universal,li2020trans,lin2020trans,schmitz2020trans,berera2020role}. A particularly interesting observation was made in \cite{brahma2020trans1} , which showed that the TCC can be derived from the Distance conjecture (1) considering that criterion to be true. The swampland conjectures have also had some pretty interesting implications on the paradigm of eternal inflation \cite{matsui2019eternal,dimopoulos2018steep,kinney2019eternal,brahma2019stochastic,wang2019eternal,trivedi2021rejuvenating}.
	\\
	\\
	The revelations from these works seemingly show that Inflation and the Swampland conjectures are in unavoidable loggerheads with each other. However, the true scenario is that it is specifically Single Field Supercooled Inflation in GR Based Cosmology, which has serious issues with the swampland criterion. Indeed, it has been shown that Inflation, even for single field cases, can be quite consistent with the swampland conjectures in other regimes of Inflation. One can take the example of warm inflation \cite{berera1995warm}, which is on amicable terms with these conjectures even for single field cases in a GR based cosmology \cite{berera2019trans,motaharfar2019warm,das2019warm,das2020swampland,das2019note,das2020distance,motaharfar1810warm,mohammadi2020warm}. Single field inflation in other cosmological scenarios which are motivated by vastly different reasons, like string theory motivated Braneworld Models \cite{randall1999alternative,randall1999large,gogberashvili2002hierarchy} or regimes like Gauss-Bonet Cosmology \cite{li2007cosmology} which are motivated by geometric requirements, have been aptly studied in the context of the swampland conjectures and have proved to be on good terms with these criterion as well \cite{lin2019chaotic,kamali2020warm,odintsov2020swampland,yi2019gauss,adhikari2020power,blumenhagen2017swampland}. Particularly in \cite{trivedi2020swampland} , it was shown that single field (cold) Inflation in a large class of non-GR based cosmologies can be quite easily consistent with the swampland conjectures. For example, this work also tangentially showed that single field inflation in loop quantum cosmology \cite{ashtekar2010loop,ashtekar2006quantum,ashtekar2006quantum1} and Chern-Simons Cosmology \cite{gomez2011standard} can also be on good terms with the swampland conjectures. Finally, it has also been shown that perhaps the Inflationary scenarios which fits the swampland conjectures the best are multi field models \cite{bravo2020tip,aragam2020multi}.
	\\
	\\
	The implications of the swampland conjectures on the current dark energy epoch of the universe are quite interesting as well. Particularly the dS conjecture (2) is in stark opposition of a cosmological constant form of dark energy and favours quintessence to be a better realization of dark energy \cite{agrawal2018cosmological,cicoli2020out,ibe2019quintessence,storm2020swampland,marsh2019swampland,olguin2019runaway,heisenberg2018dark,van2019dark,emelin2019axion,brahma2019dark,odintsov2019finite}. It was however shown in \cite{akrami2019landscape}, that all the existing string theory based models of quintessence with an exponential potential or combinations of exponential potentials are ruled out by current observations, implying that the swampland de Sitter conjecture is in tension with viable quintessence models, similarly to the single-field slow-roll inflation in a GR based cosmological scenario. The findings of that paper were then further confirmed in \cite{raveri2019swampland} with even more data. Furthermore, a significant amount of interest has also gone towards investigating non conventional quintessence models in the view of the swampland conjectures, with both multi-field and modified cosmological models having been studied in this regard \cite{akrami2020multi,cicoli2020out,oikonomou2020rescaled}. There has also been some very insightful work on studying quintessence based dark energy models in the view of the Hubble Tension \cite{Riess:2016jrr,Alam:2016hwk,mortsell2018does,vagnozzi2020new,di2020interacting,di2020nonminimal,visinelli2019revisiting} as a litmus test for the swampland conjectures in \cite{colgain2019testing,banerjee2020hubble}. Hence, there are quite a few reasons to consider quintessential inflation as a viable paradigm of dark energy and more so in the context of the swampland conjectures. If it is indeed the case for the current epoch of the universe, one might expect that the equation of state parameter of the universe can be quite different from the usual -1 in the case of a dS universe(and so a cosmological constant form of dark energy). It has, however, been shown to not quite be the case. Considering the universe went through (cold) single filed inflation in it's very early epoch and taking into account the swampland conjectures (1-3), it was shown in \cite{chiang2019does} that equation of state parameter can only diverge quite negligibly from -1. The point to note here is that this work focused on a GR based cosmology and the work done on single field inflation based in a GR based cosmology \cite{kinney2019zoo}. We have however showed in a recent paper \cite{trivedi2020swampland} that swampland conjectures have pretty different implications on single field inflation in a general class of modified cosmologies. Hence, in this paper we will be building on that work and discuss the implications that the Distance and the dS conjectures have on quintessence models in modified cosmological scenarios, given that the early universe expansion was driven by (cold) single field inflation.
	In the next section, we will be shortly reviewing the important aspects of the work discussed in \cite{trivedi2020swampland}. In Section III, we will discuss the implications of that on the equation of state parameter of the universe and see how one can get a viable distinction between constant and non-constant dark energy models through the equation of state parameter in the view of the swampland conjectures, without even having to rely to a step-function approach to inflation . We will then conclude this work in section IV with some general remarks.
	
	\section{Single field inflation in general cosmologies and the swampland conjectures}
	The central points of conflict which have been raised between single field Inflation and the primary swampland conjectures were explored deeply in \cite{kinney2019zoo}. The most serious issues which arose between these two paradigms were based on the string theory-motivated definitions of the swampland conjectures, being in grave conflict with the available observational data on single field inflation. We will quickly go over these issues now. One of the central quantities which one encounters while studying single field inflation are the slow roll parameters. A particularly important slow roll parameter is the $\epsilon$ parameter, which can be written in $m_{p} = 1 $ units in a GR based cosmology as, \begin{equation}
	\epsilon = \frac{1}{2} \left( \frac{V^{\prime}}{V} \right)^{2}
	\end{equation}
	During inflation one needs $ \epsilon << 1 $ \cite{baumann2009tasi}, and hence, it's immediately evident that the smallness requirement of the $\epsilon$ parameter is in direct conflict with the dS conjecture (2). Another issue concerns the e-fold number during Inflation, which is given the GR paradigm roughly as \begin{equation}
	N \approx \frac{ V \Delta \phi }{V^{\prime}}
	\end{equation}
	The e-fold number is a good measure of telling how much Inflation occurs in some certain model. For inflation to solve the problems of the standard big bang cosmology, one needs at least around 50 to 60 e-folds of inflationary expansion \cite{akrami2020planck1}. But it can again be immediately clear, that if one seriously considers the distance and dS conjectures (1-2), then the e-fold number would not even be greater than unity. This is an incredible assertion and one which would raise questions over the existence of any kind of single field inflationary expansion in the context of the swampland conjectures. Further in \cite{kinney2019zoo}, the following relation between the scalar spectral index \cite{aghanim2018planck,akrami2020planck1}  and the c and $ c^{\prime} $ parameters from the swampland conjectures (1-2) , was derived \begin{equation}
	1-n_{s} = [3c^{2} ; 2c^{\prime}]
	\end{equation}
	Using the data on Inflation \cite{akrami2020planck1}, one can hence deduce that $ c\leq \mathcal{O} (0.1) $ and $ c^{\prime} \leq \mathcal{O} (0.01) $, which is clearly very different then the $ \mathcal{O} (1) $ estimates of these parameters in their string theoretic form (2-3). The above equation (6), had it's roots in the following relations between the slow roll and swampland conjecture parameters, \begin{equation}
	\epsilon \geq \frac{c^{2}}{2}
	\end{equation}
	\begin{equation}
	\eta \leq - c^{\prime}
	\end{equation}
	This clearly shows that the swampland conjectures in their string theoretic form are not consistent with the data on single field inflation, in a general relativistic cosmology. While there have been works which have dealt with the swampland conjectures in some particular cosmologies \cite{lin2019chaotic,odintsov2020swampland,yi2019gauss,blumenhagen2017swampland}, in \cite{trivedi2020swampland} it was shown using a very general approach that the swampland conjectures will not have any unavoidable conflicts with single field inflation in a large class of non GR cosmological scenarios. The general Friedmann equation we considered in \cite{trivedi2020swampland}, is given by \begin{equation}
	F(H) = \frac{\rho}{3}
	\end{equation}
	where we are working in $ m_{p} = 1 $ units. Here F(H) is some function of the Hubble Parameter and $\rho$ is the energy density of the universe. Friedmann equations of this form have been used to study Inflation in detail in the paradigm of the Exact Solution Approach \cite{del2012approach,trivedi2020exact}, which is a generalization of the Hamilton-Jacobi Approach to Inflation to general cosmologies \cite{kinney1997hamilton}. Using this method, we can revaluate whether the issues which persist between single field inflation and the swampland conjectures in a GR based cosmology still hold tight in large class of cosmologies described by the general friedmann equation (9) (for more details on how general our friedmann equation in consideration is, please refer to \cite{del2012approach,trivedi2020exact}). The work in \cite{trivedi2020swampland} showed that the main issues of conflict between single field inflation and swampland conjectures highlighted in \cite{kinney2019zoo} , only necessarily hold good in a GR based cosmology and are not sources of unavoidable loggerheads between these two paradigms in the more general cosmologies described by the Friedmann equation above. 
	\\
	The $\epsilon$ parameter, for example, can be as small as is needed for a period of inflationary expansion as in general the $\epsilon$ parameter can be written as \cite{trivedi2020swampland}, \begin{equation}
	\epsilon = \frac{F H^{\prime } k}{H^{3}}
	\end{equation}
	where $ k = \frac{V^{\prime}}{V} $ and considering the dS conjecture (2) , $ k \geq c \sim \mathcal{O} (1) $. During inflation, one needs $ \epsilon << 1 $ and we can immediately start to see that in general cosmological scenarios ($ F(H) \neq H^{2} $ ), one can easily have the $\epsilon$ parameter as small as needed whilst still satisfying the dS conjecture requirement $ k > \mathcal{O} (1) $. There is no severe issue of an insufficient e-fold number as the e-fold number in this paradigm is given by , \begin{equation}
	N \approx \frac{H^{2}}{F} \frac{\Delta \phi}{k}
	\end{equation}
	where again for non-GR cosmologies, the e-fold number can satisfy both the distance and dS conjectures (1-2), and still be high enough for inflation to solve the fine tuning problems of standard big bang cosmology. And finally, the scalar spectral index is related to the c and $ c^{\prime} $ parameters as, \begin{equation}
	\left(\frac{(1 - n_{s} ) H^{2}}{2F} \right) \bigg[ \frac{1}{\left( 3 - H \frac{F_{_{\prime} H H}}{F_{_{\prime} H}} \right) \frac{H^{\prime}}{H} - \frac{H^{\prime \prime}}{H^{\prime}}} \bigg] =  \frac{F_{_{\prime} H} H^{\prime}}{F} \geq c
	\end{equation}
	\begin{equation}
	\frac{1}{F_{_{\prime} H}} \bigg[ F_{_{\prime} H H} H^{\prime} + \eta \frac{F_{_{\prime} H} H^{2}}{H^{\prime}} \bigg] \leq - c^{\prime}
	\end{equation}
	Again, the above equations (12-13) show that the $ n_{s} $ can have it's appropriate observational value \cite{aghanim2018planck,akrami2020planck1} whilst still considering the string theoretic $ \mathcal{O} (1) $ estimates of the c and $ c^{\prime} $ parameters. For the full discussion on how single field inflation and swampland conjectures can be in a peaceful harmony in general cosmological scenarios in the scope of our considered Friedmann equation (9), please refer to \cite{trivedi2020swampland}.
	\\
	\\
	\section{Swampland conjecture implications on quintessence dark energy equation of state in general cosmological scenarios }
	
	As mentioned earlier, there has been quite a lot of work on Quintessence models in the light of the swampland criterion. The swampland conjectures support the idea that the current dark energy epoch of our universe is due to rolling quintessence field rather than a positive cosmological constant \cite{agrawal2018cosmological}. It was also recently shown that considering the swampland conjectures to be true, one would not observe the equation of state parameter of the universe to be greatly different than it's dS value of -1 \cite{chiang2019does}. This revelation was made considering a GR based cosmology and as we are talking about only a single quintessence field, this result was also made on the consideration that the early universe expansion occured through single field inflation.
	\\
	We will now,however, like to show that the equation of state parameter in general cosmological scenarios described by (9), can indeed have a signficantly different value from it's dS scenario even after considering the swampland conjectures. By definition, the equation of state parameter is given by \begin{equation}
	w = \frac{p}{\rho}
	\end{equation}
	where p and $\rho$ are the pressure and energy densities of the quintessence field, respectively, given by $ p = \frac{\dot{\phi}^{2}}{2} - V(\phi) $ and $ \rho = \frac{\dot{\phi}^{2}}{2} + V(\phi) $ . Hence, \begin{equation}
	1 + w = \frac{2 \dot{\phi}^{2}}{\dot{\phi}^{2} +2 V }
	\end{equation}
	For a rolling quintessence field, we have \cite{tsujikawa2013quintessence} \begin{equation}
	3H \dot{\phi} \approx - V^{\prime}(\phi)
	\end{equation}
	which allows us to express (15) as \begin{equation}
	1 + w = \frac{2 {V^{\prime}}^{2}}{{V^{\prime}}^{2} + 18 H^{2} V}
	\end{equation}
	Further, for a rolling field in a general cosmological scenario our Friedmann equation (9) becomes \begin{equation}
	F \approx \frac{V}{3}
	\end{equation} 
	and this allows us to further rewrite (17) as \begin{equation}
	1 + w  = \frac{2 {V^{\prime}}^{2} F}{{F V^{\prime}}^{2} + 6 V^{2} H^{2}}
	\end{equation}
	Finally, considering the dS conjecture (2) we can write the above equation as \begin{equation}
	1 + w \ge \frac{2 c^{2} F}{F c^{2} + 6 H^{2}} 
	\end{equation}
	\\
	One can now see in the above inequality that for general non-GR based cosmologies ($ F \ne H^{2} $ ), w can have quite a different value than it's usual dS case even when we consider the dS conjecture. As one considers general cosmological scenarios, the parameter c can retain it's $ \mathcal{O} (1) $ estimate and still be consistent with data on single field inflation. This makes a pivotal difference, as similar analysis for the quintessence field in a GR based cosmology (with $ c \approx 0.02 $ \cite{kinney2019zoo}) resulted in a quite negligible deviation from -1 for w \cite{chiang2019does}. However, the c parameter retaining it's string theoretic value coupled with the fact that one can work with a range of different F functions corresponding to different cosmologies (again, we recommend \cite{del2012approach,trivedi2020exact} for an overview of the generality of the Friedmann equation in consideration), which would then give one a range of model-specific free parameters which could then be tuned to get quite a significant deviation from the dS value of w. It's important to also note that in the Hamilton-Jacobi approach one starts with an ansatz for the Hubble Parameter \cite{kinney1997hamilton} (contrary to the usual approach to inflationary studies where one starts off with a potential and gets the physically interesting results from that) , and in the generalizations of the Hamilton-Jacobi to general cosmologies \cite{del2012approach,trivedi2020exact} one also takes some predefined F to ascertain the cosmology they want to study inflation in. Hence, the free parameters for the model specific cases generate from these ansatz that one starts off with before making concrete calculations. This shows that in general cosmological scenarios one can have an appropriately observable distinction between constant and non constant dark energy models, whilst still considering the swampland conjectures to hold true in their string theoretic forms. This results in an exciting prospect for experiments which are aiming to measure w to within a percent's accuracy, like DES \cite{dark2005dark}, DESI \cite{flaugher2014dark} , PFS \cite{takada2014extragalactic} , Euclid \cite{laureijs2010euclid} , HSS \cite{miyazaki2012hyper} , LSST \cite{lsst2012large} and WFIRST \cite{spergel2015wide} . 
	\\
	\\
	In \cite{chiang2019does}, a step function approach towards inflation was taken and both the forms of the dS conjectures were applied in different stages of the inflationary regime owing to the tensions of the dS conjectures individually with single field inflation and the consequent negligible deviation that the w parameter has for a quintessence scenario when these conjectures are considered in a GR based cosmology. As the dS conjectures and single field inflation do not have any unavoidable conflicts with each other in general cosmological scenarios \cite{trivedi2020swampland}, one does not need adopt the step function approach to inflation in these cosmological scenarios and can have quite a different value of the w parameter from it's positive cosmological constant case. 
	\\
	\\
	\section{Concluding Remarks and Discussion }
	In this paper we have discussed the status of dark energy quintessence models in general cosmological scenarios whilst considering the swampland conjectures. We started by discussing the core issues of conflict between single fied inflation and swampland conjectues in a GR based cosmology and what these issues have implied for quintessence models. Then we briefly discussed how there are no such unavoidable conflicts between single field inflation and swampland conjectures in general cosmological scenarios, which are in the scope of a particular friedmann equation (9) and where the inflaton ( and consequently the quintessence field ) follows a usual Klein-Gordon form . We then built upon that, and showed that the equation of state parameter in general cosmological scenarios can be significantly different from it's cosmological constant value of -1 and hence, one can make a clear distinction between constant and non constant dark energy models in such cosmologies even after considering the swampland conjectures. We then concluded by noting that the absence of any serious tension between single field inflation and the conjectures in general cosmological scenarios, we hence showed that one does not need to apply any step function approach to single field inflation in order to have an equation of state parameter significantly different from -1 in the context of the swampland conjectures.

	\bibliographystyle{unsrt}
	\bibliography{VSM111.bib}
\end{document}